\documentstyle[graphicx,twocolumn,prl,aps]{revtex}
\begin{document}

\title{Commensurability oscillations due to pinned and drifting orbits in a two-dimensional lateral surface superlattice}

\author{David E. Grant and Andrew R. Long}
\address{Department of Physics and Astronomy, 
Glasgow University, Glasgow, G12~8QQ, U. K.}

\author{John H. Davies \cite{jhdaddress}}
\address{Department of Electronics and Electrical Engineering, 
Glasgow University, Glasgow, G12~8QQ, U. K.}

\date{\today}
\maketitle

\begin{abstract}
We have simulated conduction in a two-dimensional electron gas subject to a weak two-dimensional periodic potential, $V_x \cos(2\pi x/a) + V_y \cos(2\pi y/a)$. 
The usual commensurability oscillations in $\rho_{xx}(B)$ are seen with $V_x$ alone. 
An increase of $V_y$ {\em suppresses} these oscillations, rather than introducing the additional oscillations in $\rho_{yy}(B)$ expected from previous perturbation theories. 
We show that this behavior arises from drift of the guiding center of cyclotron motion along contours of an effective potential. 
Periodic modulation in the magnetic field can be treated in the same way. 
\end{abstract}

\narrowtext

The behavior of electrons in a periodic potential lies at the heart of solid state physics and continues to yield surprises. 
Motion in a {\em controllable} one (1D) or two-dimensional (2D) potential can be studied with a lateral surface superlattice (LSSL). 
The electrons typically lie in a high-mobility 2D gas in a semiconducting heterostructure, and the periodic potential is applied through an array of metal gates whose bias can be varied. 
Alternatively a patterned stressor may be used, in which case the dominant potential is piezoelectric; 
this has a lower symmetry than the stressor, which will prove important. 

The aim of using LSSLs is often to explore quantum-mechanical effects, such as Bloch oscillation and the Hofstadter butterfly, but the period of the potential is too long in most current devices. 
Instead, the dominant effects seen in 1D LSSLs are commensurability oscillations (COs) in the magnetoresistance \cite{dw89a}. 
These can be explained semiclassically \cite{cwb89a} from interference between cyclotron motion and the periodic potential. 
Consider a sinusoidal potential energy, $V(x) = V_x \cos(2\pi x/a)$. 
The interference causes a drift along the equipotentials, parallel to the $y$ axis, which contributes to the conductivity $\sigma_{yy}$ and the resistivity $\rho_{xx}$: 
\begin{equation}
{\Delta\rho_{xx}^{\rm (1D)}(V_x) \over \rho_0} = 
\left(\pi l \over a\right)^2 
\left(V_x \over E_{\rm F}\right)^2
J_0^2 \left(2\pi R_c \over a\right) . 
\label{been}
\end{equation}
Here $J_0$ is a Bessel function of the first kind, $\rho_0$ is the resistivity at $B = 0$, $l$ is the mean free path, $R_c = v_{\rm F}/\omega_{\rm c}$ is the cyclotron radius, $\omega_{\rm c} = eB/m$ is the cyclotron frequency, $E_{\rm F}$ is the Fermi energy and $v_{\rm F}$ the Fermi velocity. 
No effect on $\rho_{yy}$ is expected in this approach. 
Quantum-mechanical analysis \cite{pv89a,cz90a} yields a similar result but with small contributions to $\rho_{yy}$. 
Overall agreement between theory and experiments on 1D LSSLs is excellent, even for the strong piezoelectric potentials in strained LSSLs \cite{es96cje98,rjl98a}. 

Now consider a simple 2D potential energy, 
\begin{equation}
V(x, y) = V_x \cos(2\pi x/a) + V_y \cos(2\pi y/a) . 
\label{2D pot}
\end{equation}
To avoid ambiguity we take $V_x \ge V_y$. 
An extension of the semiclassical theory \cite{rrg92} shows that $V_x$ continues to generate oscillations in $\rho_{xx}(B)$ according to Eq.\ (\ref{been}), and $V_y$ has the same effect on $\rho_{yy}(B)$. 
In contrast to 1D potentials, there is little confirmation of this plausible result. 
An early experiment \cite{rrg91a} used a holographic technique \cite{dw89a} in two steps. 
A 1D grating was produced first, and showed strong COs in the longitudinal resistivity as expected. 
The sample was next illuminated with an orthogonal 1D pattern to produce a 2D grid. 
However, this combined pattern did {\em not} produce similar, strong COs in both $\rho_{xx}$ and $\rho_{yy}$, as expected from the extension to the semiclassical model; 
much weaker oscillations with the opposite phase were seen instead. 
Many subsequent measurements have used different modulation techniques. 
Virtually the only common feature between them is that the COs are generally weak, confirming this most significant feature of the holographic experiment. 

We have performed simulations of conduction in a 2D LSSL to address this issue, and find that $V_x$ and $V_y$ do not contribute independently. 
Instead, the introduction of $V_y$ {\em suppresses} the oscillations in $\rho_{xx}$ rather than inducing oscillations in $\rho_{yy}$. 
We explain this with a simple picture based on drift of the guiding center of cyclotron motion along contours of an effective potential. 
Trajectories can drift or be pinned, and the latter suppress the magnetoresistance. 
Remarkable behavior is found when higher Fourier components dominate this potential, which should be detectable in experiments. 

\begin{figure}
\includegraphics[scale=0.5]{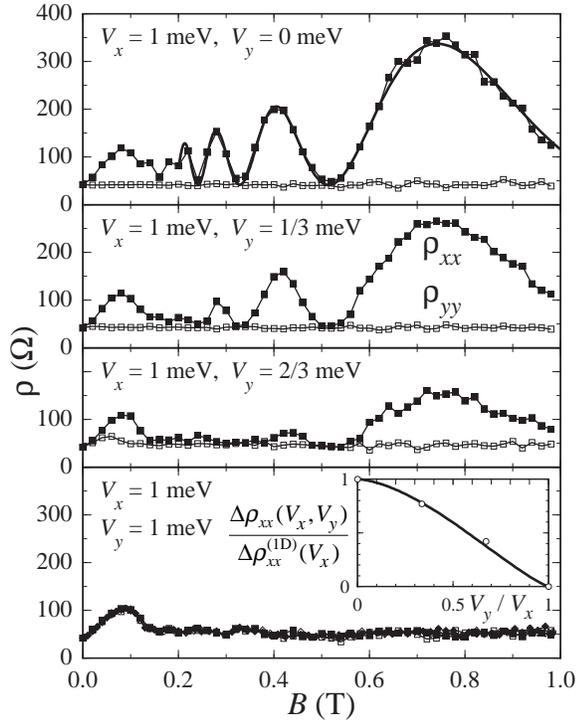}
\caption[Simulations.]
{Simulation of diagonal components of resistivity tensor for square two-dimensional superlattices with period 200\,nm, $V_x = 1\,\rm meV$, and several values of $V_y$. 
Calculated points are joined by lines for clarity and the thick curve shows the semiclassical result (Eq.\ \ref{been}) for $V_y = 0$. 
Two simulations are plotted for $V_y = V_x$ and lie on top of each other. 
The inset curve shows the estimate (Eq.\ \ref{reduced COs}) of the effect of $V_y$ on the oscillations due to $V_x$ alone, with points from the simulations.}
\label{simulation plot}
\end{figure}

To simulate conduction we solved the classical equations of motion for electrons moving in the potential energy given by Eq.\ (\ref{2D pot}) and a normal magnetic field, with a constant probability of isotropic scattering. 
The superlattice had period $a = 200\,\rm nm$ in GaAs with $3 \times 10^{15}\,{\rm m}^{-2}$ electrons of mobility $50\,{\rm m}^2\,{\rm V}^{-1}\,{\rm s}^{-1}$. 
The resistivity tensor was deduced from the velocity autocorrelation function and its diagonal elements are plotted as a function of the magnetic field in Fig.\ \ref{simulation plot}. 
We held $V_x = 1\,\rm meV$ and raised $V_y$ from zero to $V_x$. 
In the 1D limit, $V_y = 0$, the usual oscillations are seen in $\rho_{xx}$ with no structure in $\rho_{yy}$, in excellent agreement with Eq.\ (\ref{been}). 
Recall that the existing theory predicts that an increase of $V_y$ from zero should induce oscillations in $\rho_{yy}$ without affecting $\rho_{xx}$. 
Instead, we see no oscillations in $\rho_{yy}$ while those in $\rho_{xx}$ are suppressed. 
Most strikingly, there are no commensurability oscillations at all in the symmetric 2D limit where $V_y = V_x$. 
The only structure that remains is a positive magnetoresistance at low fields. 
This arises from magnetic breakdown \cite{ps90a,pb91} and it has recently been suggested \cite{ca99a} that successive breakdowns may lead to further oscillations of quantum-mechanical origin. 

\begin{figure}
\includegraphics[scale=0.5]{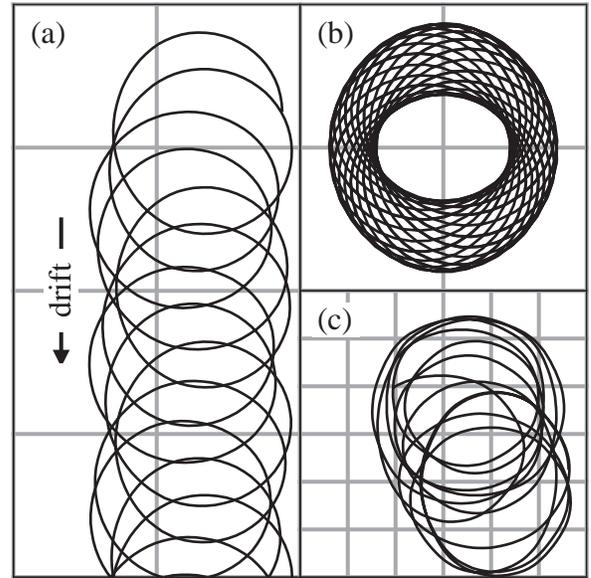}
\caption[Orbits.]
{(a)~Drifting and (b)~pinned trajectories in a square superlattice with period 200\,nm shown by the grid, $V_x = 1\,\rm meV$, $V_y = {1 \over 2}\,\rm meV$, and $B = 0.72\,\rm T$. 
(c)~Chaotic trajectory for $B = 0.28\,\rm T$; 
note the different scale.}
\label{orbits plot}
\end{figure}

An explanation of this behavior follows from the trajectories taken by electrons in the simulation. 
Typical examples starting from different points are shown in Figs.\ \ref{orbits plot}(a) and (b) for $V_x = 1\,\rm meV$ and $V_y = {1 \over 2}\,\rm meV$. 
The magnetic field $B = 0.72\,\rm T$, corresponding to the largest peak in Fig.\ \ref{simulation plot}. 
There is no scattering and the trajectories run for $100\,\rm ps$, considerably longer than the lifetime $\tau = 19\,\rm ps$ if scattering had been included. 
There is no sign of the chaos seen in weaker magnetic fields \cite{tg90a}. 
The underlying motion is clearly a cyclotron orbit and the overall trajectories can be divided into two classes. 
Fig.\ \ref{orbits plot}(a) shows the cyclotron orbit drifting along $y$. 
This is perpendicular to the wavevector of the stronger potential component, and is the only type of trajectory seen in the 1D limit, $V_y = 0$. 
Such motion contributes to $\sigma_{yy}$ and $\rho_{xx}$. 
No electrons were found to drift along $x$ and we therefore expect no effect on $\sigma_{xx}$ and $\rho_{yy}$. 
Trajectories of the second class are pinned, as in Fig.\ \ref{orbits plot}(b). 
The cyclotron orbit is distorted and precesses but makes no overall displacement. 
Such orbits make no contribution to conduction in the limit of large $\omega_{\rm c}\tau$ and therefore suppress the magnetoresistance. 
All trajectories become pinned in the symmetric 2D limit, $V_y = V_x$, quenching the COs.  
This analysis of the trajectories therefore shows that COs are reduced in magnitude because of pinned orbits, and are seen only in $\rho_{xx}$ if $V_x > V_y$. 

\begin{figure}
\includegraphics[scale=0.5]{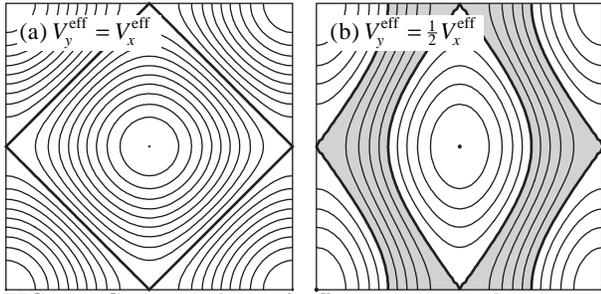}
\caption[Effective potential.]
{Contour plots of effective potential energy for a square superlattice. 
(a)~$V_y^{\rm eff} = V_x^{\rm eff}$, with all contours closed.
(b)~$V_y^{\rm eff} = {1 \over 2} V_x^{\rm eff}$, showing two bands of open contours (shaded) running parallel to the $Y$ axis.}
\label{effective potential plot}
\end{figure}

The trajectories are only weakly distorted from regular cyclotron motion. 
We therefore focus on motion of the guiding center \cite{cwb89a}, which drifts at a velocity given by 
\begin{equation}
{\bf v}^{\rm (d)}(X,Y) = \nabla V^{\rm eff}(X,Y) \times {\bf B} / eB^2 . 
\label{guiding eqn}
\end{equation}
The effective potential energy $V^{\rm eff}(X,Y)$ is the periodic potential energy (Eq.\ \ref{2D pot}) averaged over the perimeter of a cyclotron orbit centered on $(X,Y)$, which reduces $V_x$ and $V_y$ equally by a factor of $J_0(2\pi R_{\rm c}/a)$. 
This depends on magnetic field through the cyclotron radius $R_{\rm c}$. 
Eq.\ (\ref{guiding eqn}) shows that the guiding center drifts along contours of $V^{\rm eff}(X,Y)$. 
Two examples are plotted in Fig.\ \ref{effective potential plot}. 
All contours are closed for a symmetric effective potential with $V_x^{\rm eff} = V_y^{\rm eff}$ [Fig.\ \ref{effective potential plot}(a)]. 
All trajectories are therefore pinned as in Fig.\ \ref{orbits plot}(b). 
Fig.\ \ref{effective potential plot}(b) shows the effect of breaking the symmetry with $V_y^{\rm eff} = {1 \over 2} V_x^{\rm eff}$. 
This introduces a band of open contours, shaded in the plot, running parallel to the $Y$ axis. 
The guiding center can drift along these and the deviations of the contours from straight lines leads to the lateral oscillations seen in Fig.\ \ref{orbits plot}(a). 

The change in conductivity can be estimated \cite{rrg92,rrg96} from 
\begin{equation}
\Delta \sigma_{\mu\nu} = 
{e^2 m \tau \over \pi \hbar^2} 
\langle \bar v_\mu^{\rm (d)} \bar v_\nu^{\rm (d)}\rangle_{\rm orbits} \, . 
\label{delta diff}
\end{equation}
It is assumed that only drifting orbits need be considered. 
We first find the average drift velocity $\bar {\bf v}^{\rm (d)}$ for each orbit, which leaves only $\bar v_y^{\rm (d)}$ in our case. 
The square $[\bar v_y^{\rm (d)}]^2$ is then averaged over all orbits in the unit cell to give $\Delta\sigma_{yy}$. 
A difficulty is that Eq.\ (\ref{delta diff}) is valid only if the lifetime $\tau$ is much larger than the periods of the drifting orbits. 
This always fails for trajectories on the boundary of the open region because they go through stagnation points in the middle of each edge of the unit cell, but is easier to satisfy for the majority of orbits. 

The open orbits are complicated and we therefore make several approximations to estimate Eq.\ (\ref{delta diff}). 
Start from the 1D limit, $V_y = 0$, in which case all orbits drift with $\bar v_y^{\rm (d)}(X) = (2\pi V_x^{\rm eff} / eBa) \sin (2\pi X/a)$. 
The introduction of $V_y$ affects this in two ways. 
First, the fraction of the unit cell occupied by drifting orbits is reduced to 
\begin{equation}
P_{\rm drift} = 1 - {8 \over \pi^2} \int_0^{\pi/2} 
\arcsin\left(\sqrt{V_y \over V_x} \sin\theta\right) d\theta . 
\end{equation}
We replace the areas of drifting orbits shown in Fig.\ \ref{effective potential plot}(b) by bands along $Y$ of the same area centered on $X = {1 \over 4}a$ and ${3 \over 4}a$, and average $[\bar v_y^{\rm (d)}(X)]^2$ over the remaining area. 

The second effect of $V_y$ is to make the drifting orbits sinuous, which reduces their average velocity along $Y$. 
The most rapid orbit is through the symmetry point $({1 \over 4}, {1 \over 4})a$. 
Its period is increased by a factor of $(2/\pi)K$ compared with $V_y = 0$, where $K$ is the complete elliptic integral of first kind \cite{ma70} with modulus $k = V_y/V_x$. 
We apply this factor to all orbits. 
Combining the two effects lead to the approximate resistivity 
\begin{equation}
{\Delta\rho_{xx}(V_x, V_y) \over \Delta\rho_{xx}^{\rm (1D)}(V_x)}
\approx {\pi^2 \over 4K^2}
\left[P_{\rm drift} + {\sin(\pi P_{\rm drift}) \over \pi}\right] . 
\label{reduced COs}
\end{equation}
This is plotted in the inset to Fig.\ \ref{simulation plot}. 
It reduces correctly to the symmetric and 1D limits, and agrees well with the simulations. 

These results show that COs are much harder to observe in 2D potentials because the symmetry must be broken. 
They are also less robust in 2D. 
If $V_y = 0$ the guiding center drifts along $Y$, which does not change the potential seen by the electron. 
In 2D, however, the potential changes and the picture based on the guiding center will be valid only if its drift during one cyclotron period is much smaller than the unit cell. 
This leads to the condition $eBa \gg 2\pi(mV_x^{\rm eff})^{1/2}$ for COs to exist, which is similar to that for normal diffusion rather than chaos in a much stronger, symmetric potential \cite{tg90a}. 
Using the envelope of the Bessel function to relate $V_x^{\rm eff}$ and $V_x$ allows this to be rewritten as $v_{\rm F} (eBa)^3 \gg m (2\pi V_x)^2$. 
This becomes $B \gg 0.3\,\rm T$ for the conditions used in Fig.\ \ref{simulation plot}, which is in reasonable agreement with the simulations. 
We would also expect the COs to become more robust as $V_y$ is reduced, which is seen. 
Motion becomes chaotic when the condition on $B$ is violated and a typical trajectory is shown in Fig.\ \ref{orbits plot}(c) for $B = 0.28\,\rm T$. 

A general feature that follows from the drift of the guiding center along contours, Eq.\ (\ref{guiding eqn}), is that there can be only one average direction of drift. 
By symmetry this must be along $X$ or $Y$ if only $V_x$ and $V_y$ are present, so it is impossible to have oscillations in both $\rho_{xx}$ and $\rho_{yy}$. 
New features appear when further Fourier components are added to the potential. 
Consider the simplest `diagonal' component, $V_{1,1} \cos [2\pi(x+y)/a]$. 
The average over a cyclotron orbit gives an effective potential energy $V_{1,1}^{\rm eff} = V_{1,1} J_0(2\sqrt{2}\pi R_{\rm c}/a)$. 
This has a different dependence on magnetic field from $V_x$ and $V_y$, and $V_{1,1}$ can therefore dominate the behavior near zeroes of $J_0(2\pi R_{\rm c}/a)$ where the fundamental terms vanish. 

\begin{figure}
\includegraphics[scale=0.5]{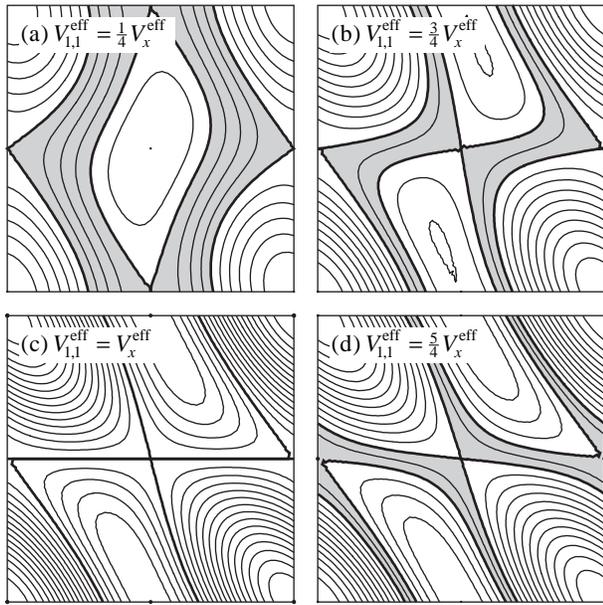}
\caption[Effect of diagonal potential.]
{Effect of diagonal component $V^{\rm eff}_{1,1}$ on contour plots of effective potential energy for a square superlattice with $V_y^{\rm eff} = {1 \over 2} V_x^{\rm eff}$. 
(a),~(b)~Region of open contours distorts as $V^{\rm eff}_{1,1}$ is raised but drift remains parallel to $Y$ on average. 
(c)~Open contours vanish for $V^{\rm eff}_{1,1} = V_x^{\rm eff}$; all orbits are pinned. 
(d)~Open contours reappear for $V^{\rm eff}_{1,1} > V_x^{\rm eff}$ but now run diagonally.}
\label{diagonal plot}
\end{figure}

The effect of raising $V_{1,1}^{\rm eff}$ from zero is displayed in Fig.\ \ref{diagonal plot}, holding $V_y^{\rm eff} = {1 \over 2} V_x^{\rm eff}$. 
The region of open contours parallel to $Y$ is distorted and shrinks as $V_{1,1}^{\rm eff}$ rises. 
It collapses when $V_{1,1}^{\rm eff} = V_x^{\rm eff}$, all orbits are pinned, and COs are quenched.
Open contours reappear when $V_{1,1}^{\rm eff} > V_x^{\rm eff}$ but are now parallel to the diagonal $Y = -X$. 
This diagonal drift induces a peak in both $\rho_{xx}$ and $\rho_{yy}$ instead of the expected minimum in $\rho_{xx}$. 
Such a mechanism may contribute to the antiphase effects seen in some experiments \cite{rrg91a}, particularly if the fundamental potential components are well balanced, and the corresponding COs are suppressed. 
Note, however, that no COs of any period will be seen within this model unless there is some asymmetry in the potentials. 

Commensurability oscillations can also be induced by a 2D periodic magnetic field, $B_z = B_0 + B_{\rm m}(x, y)$ \cite{rrg96,pv90a,pdy95c}. 
This can be analyzed in the same way with an effective potential energy given by $V^{\rm eff}_{\rm m}(X, Y) = E_{\rm F} B_{\rm m}^{\rm eff}(X, Y) / B_0$. 
Here the magnetic field $B_{\rm m}^{\rm eff}(X, Y)$ is averaged over the {\em area} of a cyclotron orbit centered on $(X,Y)$, rather than its perimeter. 
This introduces a factor of $2J_1(\theta)/\theta$ with $\theta = 2\pi R_{\rm c}/a$ for the fundamental Fourier components, rather than $J_0(\theta)$. 
Experiments show that COs in a square array of magnetic elements appear only in the direction of in-plane magnetization \cite{pdy97a} in accord with contours of $B_{\rm m}$ presented there. 

We have shown that the commensurability oscillations in a two-dimensional superlattice are quite different from the superposition of one-dimensional results. 
The addition of the potential energy $V_y \cos(2\pi y/a)$ suppresses the oscillations in $\rho_{xx}(B)$ due to $V_x \cos(2\pi x/a)$ alone, rather than adding new oscillations in $\rho_{yy}(B)$. 
There are no oscillations at all in a symmetric potential, $V_x = V_y$, provided that only open orbits contribute to conduction. 
An asymmetric potential is therefore needed to observe commensurability oscillations. 
This might seem to present difficulty, as most devices have symmetric patterns, but real structures contain strain that induces an asymmetric potential through the piezoelectric effect \cite{jhd98c}. 
The behavior in two dimensions can be explained from the drift of the guiding center of cyclotron motion along contours of an effective potential, as in one dimension, but the coupling of motion along $x$ and $y$ causes perturbation theory to fail. 
This coupling also reduces the stability of commensurability oscillations, and motion becomes chaotic at low magnetic fields. 
Higher Fourier components in the potential lead to characteristic signatures in both $\rho_{xx}$ and $\rho_{yy}$ near minima of the fundamental oscillations, and their detection would verify the theory presented here. 

It is a pleasure to thank C. J. Emeleus, B. Milton, and S. Chowdhury for many illuminating conversations. 
We also acknowledge correspondence with R. R. Gerhardts, who has similar work in progress, and the support of the U. K. EPSRC.


\end{document}